\documentclass[a4paper,12pt]{article}

\usepackage[a4paper,left=80pt,right=80pt,top=90pt,bottom=93pt]{geometry}
\usepackage[latin1]{inputenc}
\usepackage{dsfont}
\usepackage{amsfonts,amsmath,amssymb}
\usepackage{amsthm}
\usepackage{graphicx}
\usepackage[small,bf]{caption}
\usepackage{subcaption}
\usepackage{booktabs}
\usepackage[numbers, sort]{natbib}
\allowdisplaybreaks[1]
\usepackage{color}
\usepackage{ulem}
\usepackage{tikz}
\usepackage{multirow}
\usetikzlibrary{arrows}
\usepackage{xspace}

\def\bal#1\eal{\begin{align}#1\end{align}}
\newcommand{\be}{\begin{equation}}
\newcommand{\ee}{\end{equation}}
\newcommand{\bea}{\begin{eqnarray}}
\newcommand{\eea}{\end{eqnarray}}
\newcommand{\besub}{\begin{subequations}}
\newcommand{\eesub}{\end{subequations}}
\newcommand{\ba}{\begin{array}}
\newcommand{\ea}{\end{array}}
\newcommand{\bi}{\begin{itemize}}
\newcommand{\ei}{\end{itemize}}
\newcommand{\nn}{\nonumber}

\newcommand{\vev}[1]{{\ensuremath{\langle #1 \rangle}}\xspace}

\newcommand{\GeV}{{\rm GeV}\xspace}
\newcommand{\TeV}{{\rm TeV}\xspace}
\newcommand{\Mcal}{{\cal M}}

\newcommand{\Lcal}{{\cal L}}
\newcommand{\lf}{{16 \pi^2}}

\newcommand{\lh}{\ensuremath{\lambda_h}}
\newcommand{\ls}{\ensuremath{\lambda_s}}
\newcommand{\lhs}{\ensuremath{\lambda_{hs}}}

\newcommand{\cM}{{\mathcal M}}

\newcommand{\fref}[1]{Fig.~\ref{fig:#1}} 
\newcommand{\eref}[1]{Eq.~\eqref{eq:#1}}

\newcommand{\tref}[1]{Table~\ref{tab:#1}}

\begin{document}

\begin{titlepage}

\flushright{HIP-2015-8/TH}

\vspace*{1cm}

\begin{center}
{\Large 
{\bf
A second Higgs from the Higgs portal
}}
\\
[1.5cm]

{
{\bf
Adam Falkowski$^{1}$, Christian Gross$^{2}$, Oleg Lebedev$^{2}$
}}
\end{center}
%\addtocounter{footnote}{-4}

\vspace*{0.5cm}

\centering{
$^{1}$ 
\it{Laboratoire de Physique Th\'eorique, CNRS -- UMR 8627, \\
Universit\'e de Paris-Sud 11, F-91405 Orsay Cedex, France
}

\vspace*{0.15cm}
$^{2}$ 
\it{Department of Physics and Helsinki Institute of Physics, \\
Gustaf H\"allstr\"omin katu 2, FI-00014 Helsinki, Finland
}}

\vspace*{0.4cm}

\vspace*{1.2cm}

\begin{abstract}
\noindent
In the Higgs portal framework, the Higgs field generally mixes with the Standard Model (SM) singlet leading to the existence of two states, one of which is identified with the 125 GeV scalar observed at the LHC. 
In this work, we analyse direct and indirect constraints on the second mass eigenstate and the corresponding mixing angle. 
The existence of the additional scalar can be beneficial as it can stabilise the otherwise--metastable electroweak vacuum.
We find parameter regions where all of the bounds, including the stability constraints, are satisfied. We also study prospects for observing the decay of the heavier state into a pair of the 125 GeV Higgs--like scalars. 

\end{abstract}

%\today

\end{titlepage}
\newpage

\tableofcontents

%=========================================================================
%=========================================================================
\section{Introduction}
%=========================================================================
%=========================================================================

The Higgs sector of the SM has a special feature that it can couple at the renormalisable level to the hidden sector~\cite{Silveira:1985rk,Schabinger:2005ei,Englert:2013gz}. In particular, the Higgs bilinear $H^\dagger H$ is the only dimension-2 operator of the SM that is gauge and Lorentz invariant.
This allows for an interaction term
\begin{equation}
\Delta V = {\lambda_{hs}\over 2 }\, H^\dagger H \, s^2 \;,
\end{equation}
where $s$ is a real SM--singlet scalar. Given that $s$ develops a vacuum expectation value, the Higgs boson mixes with the singlet leading to the existence of two mass eigenstates $H_{1,2}$. 
In this work, we explore constraints on this scenario from various direct LEP and LHC searches, electroweak data and the Higgs couplings data.

Further motivation for exploring this model comes from stability issues of the SM. The current Higgs and top quark data favour metastability of the electroweak vacuum~\cite{Buttazzo:2013uya}.
Although the existence of a deep global minimum in the scalar potential may not be problematic for current particle physics, it does raise some questions about early Universe physics, in particular, the inflationary stage~\cite{Lebedev:2012sy}. 
These issues are avoided altogether if the Higgs potential receives a correction due to new physics which makes it convex at large field values. 
The simplest option is to couple the Higgs to a real scalar, in which case even a tiny mixing between the two can lead to a stable potential~\cite{Lebedev:2012zw,EliasMiro:2012ay}. 
Here we explore this mechanism for a more general mixing angle and study how large it is allowed to be by the current data. 
Some work in this direction has already been done in Ref.~\cite{Pruna:2013bma}, while experimental constraints on the singlet portal have also been recently discussed in Ref.~\cite{Profumo:2014opa,Chen:2014ask,Martin-Lozano:2015dja,Costa:2014qga}.
We update and extend these studies. 
We explore the full range of the singlet--like scalar masses, including the region where it is lighter than 125~GeV. 
We take into account the most up-to date constraints from coupling measurements of the 125~GeV Higgs boson, and from searches for an additional Higgs--like scalar at the LHC and other experiments. 
We also perform a comprehensive analysis of constraints on the Higgs portal scenario from electroweak precision tests.

An interesting signature of the Higgs portal is the decay of the heavier state $H_2$ into a pair of the Higgs--like states $H_1$~\cite{Englert:2011yb}, whenever it is allowed kinematically. We find, in fact, that it is allowed in most of the parameter space favoured by the stability considerations. The relevant cross section for this process is at the picobarn level for a light $H_2$, which makes it observable at the LHC run-II (see also Ref.~\cite{No:2013wsa}).

In the next section, we review the structure of the scalar potential. 
We then proceed to analysing the stability conditions, the experimental constraints and
finally implications for the LHC new physics searches.

%=========================================================================
%=========================================================================
\section{The model}
%=========================================================================
%=========================================================================

We consider an extension of the SM by a real scalar gauge-singlet field $s$, which couples to the SM Higgs field via the potential
\bal \label{pot}
V(h,s)=\frac{\lh}{4} h^4 + \frac{\lhs}{4} h^2 s^2 + \frac{\ls}{4} s^4 + \frac 12 \; \mu_h^2 \, h^2 + \frac 12 \mu_s^2 \, s^2 \,.
\eal
Here, $(0,h/\sqrt{2})$ denotes the SM Higgs doublet in the unitary gauge.
By construction, the above potential has the Z$_2$ symmetry $s \rightarrow -s$. This could also be thought of as a remnant of a U(1) symmetry in the hidden sector, under which a complex scalar field $S$ transforms and whose 
imaginary part is gauged away.

In order to produce realistic $W$ and $Z$ boson masses, $h$ must attain a VEV $\vev{h} \simeq 246.2~\GeV$.
In this paper, we consider the situation where also $s$ has a non-zero VEV.\footnote{The associated domain wall problem can be avoided either by a adding
a tiny $s^3$ term to the Lagrangian or by treating our model as a low
energy limit of a gauge theory (see above).
}
For both $h$ and $s$ non-vanishing, the potential is stationary at
\be
\vev{h}^2=\frac{2 \lhs \mu_s^2 - 4 \ls \mu_h^2}{4 \lh \ls -\lhs^2} \equiv v^2 \,,
\qquad
\vev{s}^2=\frac{2 \lhs \mu_h^2 - 4 \lh \mu_s^2}{4 \lh \ls -\lhs^2} \equiv w^2 \,.
\ee
The mass matrix at this point is
\bal
\Mcal^2= \left( \begin{array}{cc}
2 \lh v^2 & \lhs v w \\
\lhs v w & 2 \ls w^2
\end{array} \right) \,.
\eal
Since the couplings are real and we require $v^2>0,w^2>0$, the mass matrix $\Mcal^2$ is positive definite if and only if 
\begin{equation}
\lh>{\lhs^2 \over 4 \ls} ~~,~~ \ls>0 ~. \label{condition}
\end{equation}
$\Mcal^2$ can be diagonalised by the orthogonal transformation
$O^T \Mcal^2 O = \textrm{diag}(m_{H_1}^2,m_{H_2}^2) ,$
where
\bal
O&= \left( \begin{array}{cc}
\cos \theta & \sin \theta \\
- \sin \theta & \cos \theta
\end{array} \right) \;
\eal
and the angle $\theta$ satisfies
\bal
\tan 2 \theta &= \frac{ \lhs v w}{\ls w^2-\lh v^2} \,. \label{tantwotheta}
\eal
The mass squared eigenvalues are given by
\bal
m_{H_{1,2}}^2 = \lh v^2+\ls w^2\mp \frac{ \ls w^2 - \lh v^2}{\cos 2 \theta} \,.
\label{evalues}
\eal
Note that we are using a different convention for $\theta$ compared to that of~\cite{Lebedev:2011aq,Lebedev:2012zw}.
 The above equation implies $\textrm{sign}(m_{H_2}^2-m_{H_1}^2)=\textrm{sign}(\cos 2 \theta) \,\textrm{sign}(\ls w^2 - \lh v^2)$.
The fields in the mass eigenstate basis are
\bal
\left( \begin{array}{c}
H_1\\
H_2
\end{array} \right)
=
 \left( \begin{array}{c}
\cos \theta \left (h - \vev{h} \right ) - \sin \theta \left (s - \vev{s} \right )\\
\cos \theta \left (s - \vev{s} \right ) + \sin \theta \left (h - \vev{h} \right )
\end{array} \right)\,.
\eal
In the following, $H_1$ is always identified with the 125~GeV boson discovered at the LHC. 

As we are interested in stability properties of the vacuum,
it is useful to point out that our Z$_2$-symmetric potential 
 subject to (\ref{condition}) has a $single$ local minimum at tree level (barring the reflected minimum $w \rightarrow -w$). Indeed, as 
 detailed in~\cite{Lebedev:2011aq}, the stationary points are local minima under the following conditions:
\begin{eqnarray}
&& v\not=0, w\not=0~:~ \lhs \mu_s^2 - 2 \ls \mu_h^2 >0~,~ \lhs \mu_h^2 - 2 \lh \mu_s^2>0 ~, \nonumber \\
&& v\not=0, w=0~:~ \lhs \mu_h^2 - 2 \lh \mu_s^2<0 ~,~ \mu_h^2 <0 ~, \nonumber \\
&& v=0, w\not=0~:~ \lhs \mu_s^2 - 2 \ls \mu_h^2 <0~,~\mu_s^2 <0 ~, \nonumber \\
&& v=0, w=0~:~ \mu_h^2 >0~,~ \mu_s^2 >0 ~.
\end{eqnarray}
These conditions are not compatible with each other and only one of them
can correspond to a local minimum. As long as radiative corrections are small, e.g. when there are no large logs,
this situation persists at the loop level. However, at large field values additional minima may develop.

We will consider the possibility that the SM extended by the singlet is valid up to the Planck scale.
This entails constraints on the couplings $\lambda_i$ as those must remain perturbative and lead to a stable scalar potential. We will require absolute stability of the electroweak vacuum. Although metastability is
sufficient for many applications, reconciling the existence of a deeper minimum with cosmology may be 
non-trivial. For that reason, we choose to impose the stronger condition.

Electroweak scale constraints are formulated more easily in terms of the parameters
$(m_{H_1}^2, m_{H_2}^2,\sin \theta,v,\lhs)$. On the other hand, perturbativity and stability analyses favour the 
set $(\lh,\ls,\lhs,v,w)$. The quartic couplings can be expressed in terms of the ``more physical'' parameters
as 
\besub
\bal
\lh &=\frac{m_{H_1}^2}{2 v^2} + \sin^2 \theta \ \frac{m_{H_2}^2-m_{H_1}^2}{2 v^2} \label{lhLE}
\\
\ls &= \frac{ 2 \, \lhs^2}{\sin^2 2 \theta } \frac{v^2}{m_{H_2}^2-m_{H_1}^2} \left(\frac{m_{H_2}^2}{m_{H_2}^2-m_{H_1}^2} -\sin^2 \theta \right) \;. \label{lsLE}
\eal
\eesub
We leave $\lhs$ as an independent variable as it is directly related to the decay rate
of $H_2$ into a pair of $H_1$'s, when the process is allowed kinematically.

Let us now write down the couplings of the scalars to the SM matter. 
Those involving a single scalar are given by 
\be
\Lcal \supset { H_1 \cos \theta + H_2 \sin \theta \over v} \left [ 2 m_W^2 W_\mu^+ W^{\mu -} + m_Z^2 Z_\mu Z^\mu - \sum_f m_f \bar f f \right] \,.
\ee
Thus, the partial decay widths of $H_1$ into SM matter are universally suppressed with respect to those of the SM Higgs by $\cos^2\theta$. 
Similarly, the partial decay widths of $H_2$ into SM matter are those of a would-be SM Higgs with mass $m_{H_2}$ universally suppressed by $\sin^2\theta$.

 On top of that $H_2$ decays to $H_1$ pairs are possible if $m_{H_2} > 2 m_{H_1} \sim 250$~GeV, and $H_1$ decays into $H_2$ pairs are possible for $m_{H_2} < m_{H_1}/2 \sim 62.5$~GeV. 
 These decays are mediated by the scalar self-interaction terms which we parametrise as 
 \be
\Lcal \supset -{\kappa_{112} \over 2} v \sin \theta ~H_1^2 H_2 -{\kappa_{221} \over 2} v \cos \theta ~H_2^2 H_1 ~,
\ee
where the couplings are given by 
\besub
\bal
\kappa_{112} &= 
 {2 m_{H_1}^2 + m_{H_2}^2 \over v^2 } \left ( \cos^2 \theta + { \lambda_{hs} v^2 \over m_{H_2}^2 - m_{H_1}^2} \right )~,
\\ 
\kappa_{221} &= 
 {2 m_{H_2}^2 + m_{H_1}^2 \over v^2 } \left ( \sin^2 \theta + { \lambda_{hs} v^2 \over m_{H_1}^2 - m_{H_2}^2} \right ) \,.
\eal
\eesub
In the kinematically allowed regime the decay widths are given by
\besub
\bal \label{gamma211}
\Gamma(H_2 \to H_1 H_1) &= {\sin^2 \theta ~\kappa_{112}^2 v^2 \over 32 \pi \, m_{H_2}} \sqrt{1 - {4 m_{H_1}^2 \over m_{H_2}^2}}~,
\\ 
\Gamma(H_1 \to H_2 H_2) &= {\cos^2 \theta ~\kappa_{122}^2 v^2 \over 32 \pi \, m_{H_1}} \sqrt{1 - {4 m_{H_2}^2 \over m_{H_1} ^2}} \,.
\eal
\eesub

%=========================================================================
%=========================================================================
\section{Vacuum stability and perturbativity}
%=========================================================================
%=========================================================================

Here we study what constraints are imposed on the parameter space if we require the couplings to remain perturbative and the electroweak vacuum to be stable all the way up to the Planck scale (see also~\cite{Khoze:2014xha,Costa:2014qga}).

The potential has 5 parameters of which 2 are fixed by requiring $v=246.2~\GeV$ and $m_{H_1}=125.15~\GeV$.
As the remaining 3 parameters we choose the mass $m_{H_2}$ of the heavier state $H_2$, the admixture $\sin \theta$ of the singlet to the state $H_1$ and the coupling $\lhs$.
For given values of $m_{H_2},\sin \theta$ and $\lhs$, the corresponding values of $\lh$ and $\ls$ are determined by Eqs.~(\ref{lhLE},\ref{lsLE}).
The resulting couplings $\lh,\ls$ and $\lhs$ are evolved to the Planck scale $m_P=2.4 \times 10^{18}~\GeV$ using one-loop RG evolution.
The relevant RGEs (neglecting all the Yukawa couplings except for $y_t$) read 
\bal
16\pi^2 {d \lambda_h \over dt}
&= 
24 \lambda_h^2 -6 y_t^4 + \frac38
\left( 2 g^4 + (g^2 + g^{\prime 2})^2 \right) 
+ (-9 g^2 -3 g^{\prime 2}+12 y_t^2) \lambda_h + \frac12 \lambda_{hs}^2 \;, \nn
\\
16\pi^2 {d \lambda_{hs} \over dt} 
&= 
4 \lambda_{hs}^2 + 12 \lambda_h \lambda_{hs}
-\frac32 (3 g^2 + g^{\prime 2}) \lambda_{hs} 
+ 6 y_t^2 \lambda_{hs} + 6 \lambda_s \lambda_{hs} \;, \nn
\\
 16\pi^2 {d \lambda_{s} \over dt} 
 &= 
 2 \lambda_{hs}^2 + 18 \lambda_s^2 \;, \nn
\\
\lf \frac{d y_t}{dt}&= y_t \left(\frac92 y_t^2 - \frac{17}{12} g'^2 - \frac94 g^2 - 8 g_3^2 \right) \;, \nn
\\
\lf \frac{d g_i}{dt}&= b_i \, g_i^3 \quad \textrm{with} \quad (b_1,b_2,b_3)=(41/6,-19/6,-7) \;,
\eal
where $g_i=(g',g,g_3)$ denotes the gauge couplings.
As input values we use $g(m_t)=0.64,\; g'(m_t)=0.35,\; g_3(m_t)=1.16$ and $y_t(m_t)=0.93$.
Our input top Yukawa coupling is based on the central value of $m_t(m_t)$
advocated in~\cite{Moch:2014lka}.

The vacuum stability conditions depend on the sign of $\lhs$ (cf. also the discussion in Ref.~\cite{EliasMiro:2012ay}):
\paragraph{\underline{${\lhs>0}$:}}
The requirement $\lh>\lhs^2/(4 \ls)$ has to be met only around the mass scale of the fields (i.e. around the \TeV scale in our case) in order for $v,w$ to be a minimum of the potential.
It may not however hold at the high energy scale. As long as the quartic couplings are positive $\lambda_i >0$, the potential is positive definite and no 
run-away direction exists.
\paragraph{\underline{${\lhs<0}$:}}
Neglecting the quadratic terms, the potential can be written as 
\be
V \simeq \frac14 \left[ \left(\sqrt{\lh} \, h^2 - \sqrt{\ls} \,s^2\right)^2+\left(\lhs+2 \sqrt{\lh \ls}\right)h^2 s^2 \right]\,.
\ee
 This shows that $V$ has a run-away direction at large field values unless $\lh>\lhs^2/(4 \ls)$.
This condition and $\ls>0$ are therefore to be imposed at all scales.

\begin{figure}[ht] 
\includegraphics[width=0.47\textwidth]{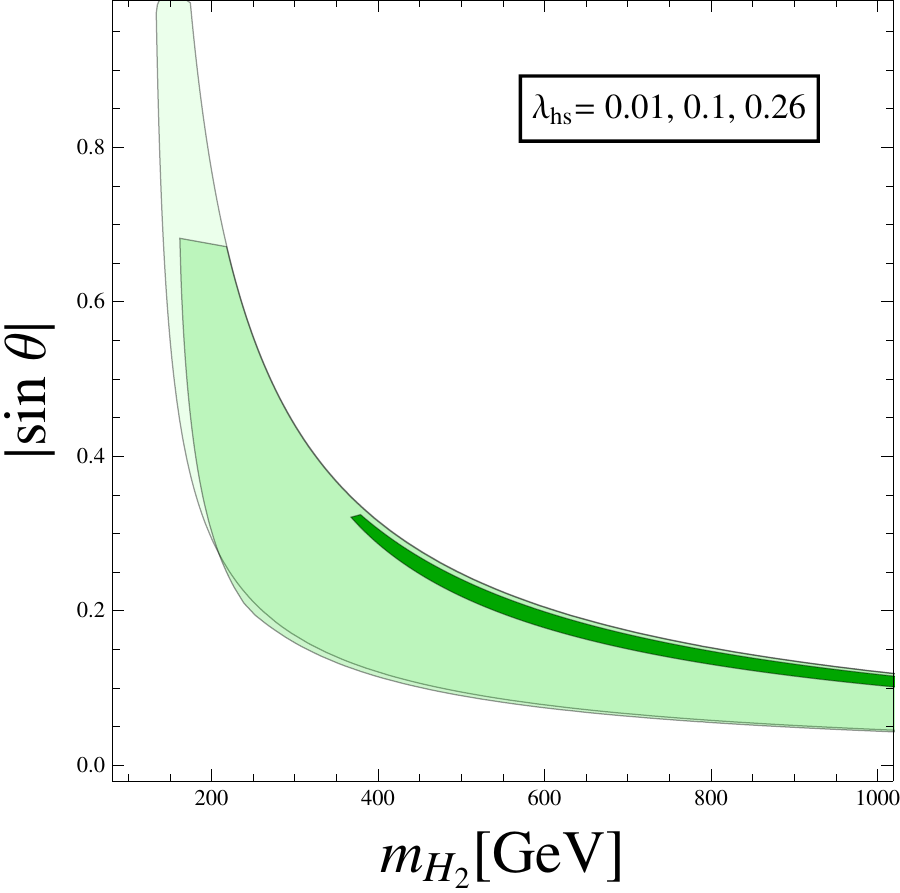}
\qquad
\includegraphics[width=0.47\textwidth]{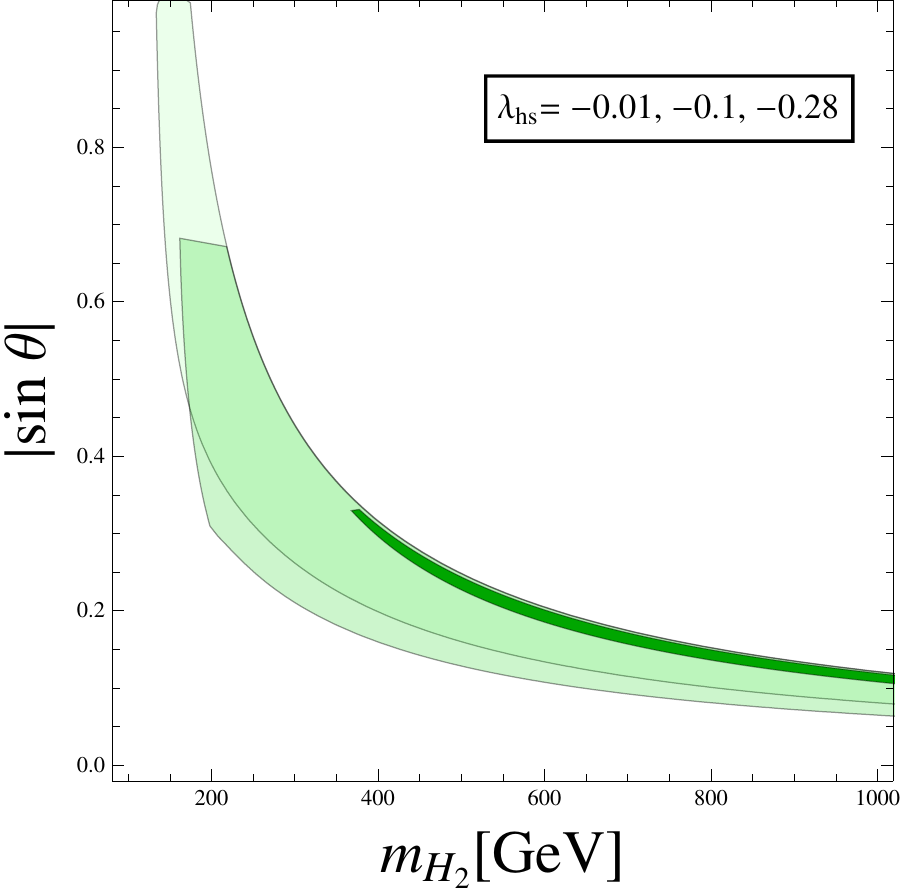}
\caption{ \label{plotPertStab}
{\em Left:} Regions of parameter space (shaded) where the couplings remain perturbative and the electroweak vacuum remains stable up to the Planck scale, for $\lhs=0.01$ at the electroweak scale (lightest green), $\lhs=0.1$ (green), $\lhs=0.26$ (darkest green).
{\em Right:} Analogous plot for negative values of $\lhs$, namely $\lhs=-0.01$ (lightest green), $\lhs=-0.1$ (green), $\lhs=-0.28$ (darkest green).
}
\end{figure}
In Figure~\ref{plotPertStab}, we show, for different weak scale values of $\lhs$, the area in the $m_{H_2} - | \sin \theta \,|$ plane where $\lh,\ls,\lhs$ remain perturbative ($\lambda_i< 4 \pi^2$) up to $m_P$ and where the electroweak vacuum remains stable.\footnote{Note that imposing a stricter criterion for perturbativity, such as $\lambda_i < 4 \pi$ or $\lambda_i < 1$, affects the allowed region in the parameter space only mildly.
This stems from the fact that an ${\cal O}(1)$ coupling becomes nonperturbative
very quickly. }
Qualitatively, the shape of the allowed regions can be understood as follows:
 \bi
 \item
At the upper boundary of each of the allowed regions the coupling $\lh$ becomes non-perturbative below $m_P$.
Note that the initial value for $\lh$ at the weak scale grows with $|\sin \theta|$ as well as with $m_{H_2}$.
Since the beta function of $\lh$ has a positive contribution proportional to $\lh^2$, it is clear that above a certain curve (which roughly has the shape $|\sin \theta| \propto 1/m_{H_2}$), the initial values are so large that $\lh$ does not remain perturbative up to $m_P$.

\item
The limiting factor that determines the lower edge of the allowed regions depends on the value of $\lhs$:
\bi
\item
For small $\lhs$ ($\lhs=\pm 0.01$), the lower edge is
determined by stability of the potential.
Since the initial value of $\lh$ decreases with decreasing $|\sin \theta|$ as well as with $m_{H_2}$, below a certain curve (which again roughly has the shape $|\sin \theta| \propto 1/m_{H_2}$), the additional threshold contribution $\lh - \lh^{SM}$ becomes too small 
to either keep $\lh>0$ (for $\lhs>0$) or satisfy
 $\lh>\lhs^2/(4 \ls)$ (for $\lhs<0$) all the way up to $m_P$.
\item

For sizeable $\lhs$ ($\lhs=0.26$ and $\lhs=-0.28$), perturbativity of $\ls$ is more constraining than stability of the vacuum, i.e. at the lower edge of the allowed regions the coupling $\ls$ becomes non-perturbative below $m_P$.
Since, for substantial $m_{H_2}$ and small $\sin\theta$,
 $\ls \propto 1/\sin^2 \theta \times 1/ m_{H_2}^2$, the lower edge has the shape $|\sin \theta| \propto 1/m_{H_2}$.

The reason why perturbativity becomes more constraining lies in the 
positive contribution $\propto \lhs^2$ to the beta-functions of $\ls$
and $\lhs$. Increasing $|\lhs|$ shrinks the allowed region further,
eventually leaving no allowed parameter space.

\item
For intermediate values of $\lhs$ ($\lhs= \pm 0.1$), the limiting factor at the upper left side of the lower edge is perturbativity of $\ls$, while vacuum stability is the limiting factor for the rest of the lower edge.
The transition between the two is at the (hardly visible) kink of the lower edge of the plots.
\ei

\ei

Finally, let us note that if $H_2$ is lighter than the SM--like state $H_1$, 
the quartic coupling $\lambda_h$ at the electroweak scale is $smaller$ than that in the SM, which makes it harder to achieve stability of the electroweak vacuum. On the other 
hand, the one--loop correction due to $\lambda_{hs}$ is positive and, if sufficiently large, could overcome the above tree--level setback. We find that this is possible
if $4 \lambda_h \lambda_s - \lambda_{hs}^2$ is positive yet very close to 
zero at $m_t$, with roughly $\lambda_{hs} \sim 0.3165$, $m_{H_2} \lesssim 0.6$ GeV and $\sin \theta \lesssim 0.04$. However, we find that $4 \lambda_h \lambda_s - \lambda_{hs}^2$ runs negative already tens of GeV above $m_t$, which shows that 
to establish the existence of this minimum in the scalar potential would require a more sophisticated analysis of the full Coleman--Weinberg potential including the 2--loop effects. Since this region of parameter space is excluded experimentally 
due to the efficient $H_1 \rightarrow H_2 H_2$ decay (cf. section~\ref{sec:indirectconstaints}), we do not study it in more detail.

%=========================================================================
%=========================================================================
\section{Experimental constraints}
%=========================================================================
%=========================================================================

In this section we discuss the experimental constraints on a singlet scalar mixing with the Higgs boson.

%=========================================================================
\subsection{Limits from electroweak precision data}
%=========================================================================

In our model, electroweak observables are affected at leading order only via oblique corrections, that is via one-loop contributions to the propagators of the W and Z bosons. 
These corrections come from two sources: i) loop diagrams with the new scalar $H_2$, and ii) modified couplings of the 125~GeV scalar $H_1$ to the gauge bosons.

We define the propagator function $\Pi_{V V}$ via the 2-point amplitude $\cM(V_{\mu} \to V_{\nu}) = \eta_{\mu\nu} \Pi_{V V}(p^2) + \dots$. 
In dimensional regularisation, the shift of the propagator function with respect to the SM is given by 
\be
\label{eq:dpivv} 
\delta \Pi_{VV}(p^2) = {m_V^2 \sin^2 \theta \over 4 \pi^2 v^2} \left [ 
{m_{H_2}^2 - m_{H_1}^2 \over 4 } \left ( {1 \over \epsilon } + 1 \right ) + F (p^2,m_V^2,m_{H_2}^2)-F (p^2,m_V^2,m_{H_1}^2)\right ], 
\ee
where $V = W,Z$ and the loop function $F$ is defined by
\be
F (p^2,m_V^2,m_\phi^2) = \!\! \int_0^1 \!\!\! dx \left [ m_V^2- {\Delta \over 2} \right ] \log \Delta, 
\quad \textrm{with} \quad
\Delta = x m_\phi^2 + (1-x) m_V^2 - p^2 x (1-x) \,.
\ee
The $1/\epsilon$ divergence cancels in physical observables. 
 \begin{table}[h!]
 \small
 \begin{center}
 \begin{tabular}{|l|l|c|l|l|}
\hline
observable & experimental value & Ref. & SM prediction & definition
 \\ \hline \hline
$\Gamma_{Z}$ [GeV] & $2.4952 \pm 0.0023$ & \cite{ALEPH:2005ab} & $ 2.4950$ & $\sum_f \Gamma (Z \to f \bar f)$ 
 \\ \hline
$\sigma_{\rm had}$ [nb] & $41.540\pm 0.037$ & \cite{ALEPH:2005ab} & $41.484$ & ${12 \pi \over m_Z^2} {\Gamma (Z \to e^+ e^-) \Gamma (Z \to q \bar q) \over \Gamma_Z^2}$ 
 \\ \hline 
 $R_{\ell}$ & $20.767\pm 0.025$ & \cite{ALEPH:2005ab} & $20.743$ & $ {\sum_{q} \Gamma(Z \to q \bar q) \over \Gamma(Z \to \ell^+ \ell^-)} $
 \\ \hline
 $A_\ell$ & $0.1499 \pm 0.0018$ & \cite{Baak:2014ora} & $0.1472$ & ${\Gamma(Z \to e_L^+ e_L^-) - \Gamma(Z \to e_R^+ e_R^-) \over \Gamma(Z \to e^+ e^-) }$ \\ \hline
 $A_{\rm FB}^{0,\ell}$ & $0.0171\pm 0.0010$ & \cite{ALEPH:2005ab} & $0.0163$ & ${3 \over 4} A_\ell^2$ 
 \\ \hline
 $\sin^2\theta_{\rm eff}^\ell(Q_{\rm FB})$ & $0.2324\pm 0.0012$ & \cite{ALEPH:2005ab} & $0.23150$ & 
 ${g_Y^2 \over g_L^2 + g_Y^2} (1 - {g_L \delta \Pi_{Z \gamma}(m_Z^2) \over g_Y m_Z^2} )$ 
 \\ \hline
$R_b$ & $0.21629\pm0.00066$ & \cite{ALEPH:2005ab} & $0.21578$ & ${ \Gamma(Z \to d \bar d) \over \sum_q \Gamma(Z \to q \bar q)}$ 
 \\ \hline
$A_b$ & $0.923\pm 0.020$ & \cite{ALEPH:2005ab} & $0.935$ &
 ${ \Gamma(Z \to d_L \bar d_L) - \Gamma(Z \to d_R \bar d_R) \over \Gamma(Z \to d \bar d) }$ 
 \\ \hline
$A_{b}^{\rm FB}$ & $0.0992\pm 0.0016$ & \cite{ALEPH:2005ab} & $0.1032$ & ${3 \over 4} A_\ell A_b$ \\ \hline
$R_c$ & $0.1721\pm0.0030$ & \cite{ALEPH:2005ab} & $0.17226$ 
& ${ \Gamma(Z \to u \bar u) \over \sum_q \Gamma(Z \to q \bar q)} $ 
\\ \hline
$A_c$ & $0.670 \pm 0.027$ & \cite{ALEPH:2005ab} & $0.668$ 
 & ${ \Gamma(Z \to u_L \bar u_L) - \Gamma(Z \to u_R \bar u_R) \over \Gamma(Z \to u \bar u) }$ 
\\ \hline 
$A_{c}^{\rm FB}$ & $0.0707\pm 0.0035$ & \cite{ALEPH:2005ab} & $0.0738$ & ${3 \over 4} A_\ell A_c$ 
\\ \hline \hline 
 $m_{W}$ [GeV] & $80.385 \pm 0.015$ & \cite{Group:2012gb} & $80.3602$ & $\sqrt{{g_L^2 v^2 \over 4} + \delta \Pi_{WW}(m_W^2)}$
\\ \hline 
$\Gamma_{W}$ [GeV] & $ 2.085 \pm 0.042$ & \cite{Beringer:1900zz} & $2.091$ & $ \sum_f \Gamma(W \to f f')$ \\
\hline
${\rm Br} (W \to {\rm had})$ & $ 0.6741 \pm 0.0027$ & \cite{Schael:2013ita} & $0.6751$ & $ {\sum_q \Gamma(W \to q q') \over \sum_f \Gamma(W \to f f')}$
 \\ \hline
\end{tabular}
\end{center}
\caption{
The electroweak precision observables used in this analysis. 
We take into account the experimental correlations between the LEP-1 Z-pole observables and between the heavy flavour observables. 
For the theoretical predictions we use the best fit SM values from GFitter~\cite{Baak:2014ora}, except for ${\rm Br} (W \to {\rm had})$ where we take the value quoted in~\cite{Schael:2013ita}. 
 }
\label{tab:EWPT_pole}
\end{table}

The observables used in our fit are the LEP-1 Z-pole observables~\cite{ALEPH:2005ab}, the W mass~\cite{Group:2012gb}, the total width~\cite{Beringer:1900zz}, and the hadronic width~\cite{Schael:2013ita}, see \tref{EWPT_pole}.
The W and Z partial decay widths appearing in the table are given by
\be
\label{eq:pw}
\Gamma(Z \to f \bar f) = {N_f m_Z \over 24 \pi} g_{f Z;\rm eff}^2, 
\qquad \Gamma(W \to f f') = {N_f m_W \over 48 \pi} g_{fW; \rm eff} ^2
\ee
where $N_f$ is the number of colours of the fermion $f$ and the effective couplings are defined as (see e.g.~\cite{Wells:2005vk})
\bal
\label{eq:EWPT_geff} 
g_{fZ; \rm eff} &=
 { \sqrt{g_L^2 + g_Y^2} \over \sqrt{ 1 - \delta \Pi_{ZZ}'(m_Z^2)}} \left [T^3_f - Q_f s_{\rm eff}^2 \right ], 
 \quad
 s_{\rm eff }^2 = {g_Y^2 \over g_L^2 + g_Y^2} \left (1 - {g_L \over g_Y} {\delta \Pi_{\gamma Z}(m_Z^2) \over m_Z^2} \right ), 
\nn \\ 
 g_{fW; \rm eff} &= g_{W; \rm eff} = {g_L \over \sqrt{ 1 - \delta \Pi_{WW}'(m_W^2)}} \,, 
\eal
where $g_L$ and $g_Y$ are the gauge couplings of $SU(2) \times U(1)$. 
Note that in our model $\delta \Pi_{\gamma Z}$ as well as $\delta \Pi_{\gamma \gamma}$ vanish at one-loop level.
As is customary, the SM electroweak parameters $g_L$, $g_Y$, $v$ are taken from the input observables $G_F, \alpha$ and $m_Z$. 
The oblique corrections also contribute to our input observables, effectively shifting the electroweak parameters by
\bal
\label{eq:EWPT_input_shift}
{\delta g_L \over g_L} &= {1 \over g_L^2 - g_Y^2} \left ( 
 2{ \delta \Pi_{WW}(0) \over v^2} 
 - 2 \cos^2 \theta_W { \delta \Pi_{ZZ}(m_Z^2) \over v^2} 
+ {g_Y^2 \over 2} \delta \Pi_{\gamma \gamma}'(0)
\right ),
\nn
\\ 
\nn
 {\delta g_Y \over g_Y} &= {1 \over g_L^2 - g_Y^2} \left ( 
- {2 g_Y^2 \over g_L^2 } {\delta \Pi_{WW}(0) \over v^2} 
+ 2 \sin^2 \theta_W { \delta \Pi_{ZZ}(m_Z^2) \over v^2} 
- {g_L^2 \over 2 } \delta \Pi_{\gamma \gamma}'(0) 
\right ),
\\ 
 {\delta v \over v} &= -{ 2 \delta \Pi_{WW}(0) \over g_L^2 v^2} \,.
\eal
Using Eqs.~(\ref{eq:EWPT_geff})~and~(\ref{eq:EWPT_input_shift}) one can calculate how the effective couplings (and hence, by Eq.~(\ref{eq:pw}), the partial decay widths) are shifted in the presence of oblique corrections, and compute the corrections to precision observables. 
We take into account the leading order (linear) corrections in $\delta \Pi_{VV}$. 
Using the observables in Table~\ref{tab:EWPT_pole}, we construct a global $\chi^2$ function that depends on $m_{H_2}$, $\sin \theta$, and known SM parameters. 
For each $m_{H_2}$ we minimise the global $\chi^2$ with respect to $\sin \theta$, and determine the 95\% CL limits by solving \be
\chi^2(m_{H_2}, \sin \theta) - \textrm{min}_{\theta} \{ \chi^2(m_{H_2}, \sin \theta)\} = 3.84\,.
\ee
The excluded region is shown as the grey area in \fref{all}. 
The limits are non-trivial for $m_{H_2} \lesssim 60$~GeV and $m_{H_2} \gtrsim 170$~GeV, and become stronger as $H_2$ gets heavier. 
For $m_{H_2} \gtrsim 450$~GeV, the electroweak precision constraints provide the strongest limits on the model.\footnote{Using Eq.~(\ref{eq:dpivv}) one could also compute the usual Peskin-Takeuchi S and T parameters, which in the case at hand are 
$S = {16 \pi \cos^2 \theta_W \over g^2} \delta \Pi_{ZZ}'(0)$, 
$T = {4 \pi \over e^2} \left ( {\delta \Pi_{WW}(0) \over m_W^2} - {\delta \Pi_{ZZ}(0) \over m_Z^2} \right )$. 
For $m_{H_2} \gg m_{H_1}$ this gives
\be
\label{eq:STapp}
T \approx - {3 \over 8 \pi \cos^2 \theta_W} \sin^2 \theta \log (m_{H_2}/M_T) , 
\qquad
S \approx {1 \over 6 \pi } \sin^2 \theta \log (m_{H_2}/M_S), \nn
% \qquad M_T = m_W^{\beta_W} m_Z^{\beta_Z} m_{H_1}^{1- \beta_W - \beta_Z} 
%\qquad M_S = e^{-\tilde \gamma_Z} m_Z^{\gamma_Z} m_{H_1}^{1- \gamma_Z}
\ee
where $M_T \approx 211$~GeV, $M_S \approx 81$~GeV. 
The resulting constraints from $S$ and $T$ indeed give a good approximation (within 10\%) of the actual limits for $m_{H_2} \gtrsim 400$~GeV. 
We stress that our analysis is valid for any $m_{H_2}$, in particular also for $m_{H_2} \ll m_{H_1}$. 
}

%=========================================================================
\subsection{Limits from Higgs coupling measurements} \label{sec:indirectconstaints}
%=========================================================================

An important constraint on the model comes from the fact that mixing with the singlet modifies the coupling strength of the Higgs boson to the SM gauge bosons and fermions. 
The couplings of the 125 GeV boson, here identified with $H_1$, have recently been measured at the LHC in several decay channels. 
Here we only use the results with the $\gamma \gamma$ and $4 \ell$ final states that have the best mass resolution. 
This allows us to simplify the discussion of possible contamination of the $H_1$ signal strength measurements by $H_2$ decays. 
We will assume that for $m_{H_2}$ outside the interval $[120, 130]$~GeV this contamination is absent, as suggested by the results of ATLAS and CMS Higgs searches in these two channels. 
In order to determine the limits on $\sin \theta $ for $m_{H_2} \in [120, 130]$~GeV, one needs a more elaborate analysis that takes into account a different mass resolution in various $h\to \gamma \gamma$ and $h\to 4 \ell$ search categories. 
We will not attempt such an analysis in this paper.
 \begin{table}[h]
 \begin{center}
 \begin{tabular}{|c|c|c|}
\hline
Channel & $\mu$ (ATLAS) & $\mu$ (CMS)
 \\ \hline \hline 
$H_{1} \to \gamma \gamma$ & $1.17^{+0.27}_{-0.27}$ \cite{Aad:2014eha} & $1.12^{+0.24}_{-0.24}$ \cite{Khachatryan:2014jba} \\ \hline 
$H_{1} \to ZZ^* \to 4 \ell$ & $1.44^{+0.40}_{-0.33}$ \cite{Aad:2014eva} & $1.00^{+0.29}_{-0.29}$ \cite{Khachatryan:2014jba} \\ \hline \hline 
\end{tabular}
\end{center}
\caption{
The signal strength of the 125~GeV scalar relative to that of the SM Higgs measured at the LHC in the $\gamma \gamma$ and $4 \ell$ channels. 
}
\label{tab:Higgs125}
 \end{table}

We use the most recent inclusive $H_1$ signal strengths measurements by ATLAS and CMS collected in \tref{Higgs125}. 
Moreover, we take into account the $15\%$ theoretical uncertainty in the Higgs production cross section, which is a linear sum of the PDF and QCD scale errors on the gluon fusion cross section~\cite{Dittmaier:2011ti}. 
We include this uncertainty as a Gaussian-modeled nuisance parameter.
With this procedure, we get the combined constraint on the Higgs signal strength 
\be
\mu > 0.81, \qquad @95\% \ {\rm CL}. 
\ee
For $m_{H_2} \geq m_{H_1}/2 \sim 62.5$~GeV and $m_{H_2}$ outside the [120,130]~GeV interval, this translates to a bound on $\sin \theta$, 
\be
\label{eq:indirect_sinth}
\sin \theta < 0.44, \qquad @95\% \ {\rm CL}, 
\ee
that is independent of $m_{H_2}$. 
\begin{figure}[h!]
 \begin{center}
 \includegraphics[width=0.47\textwidth]{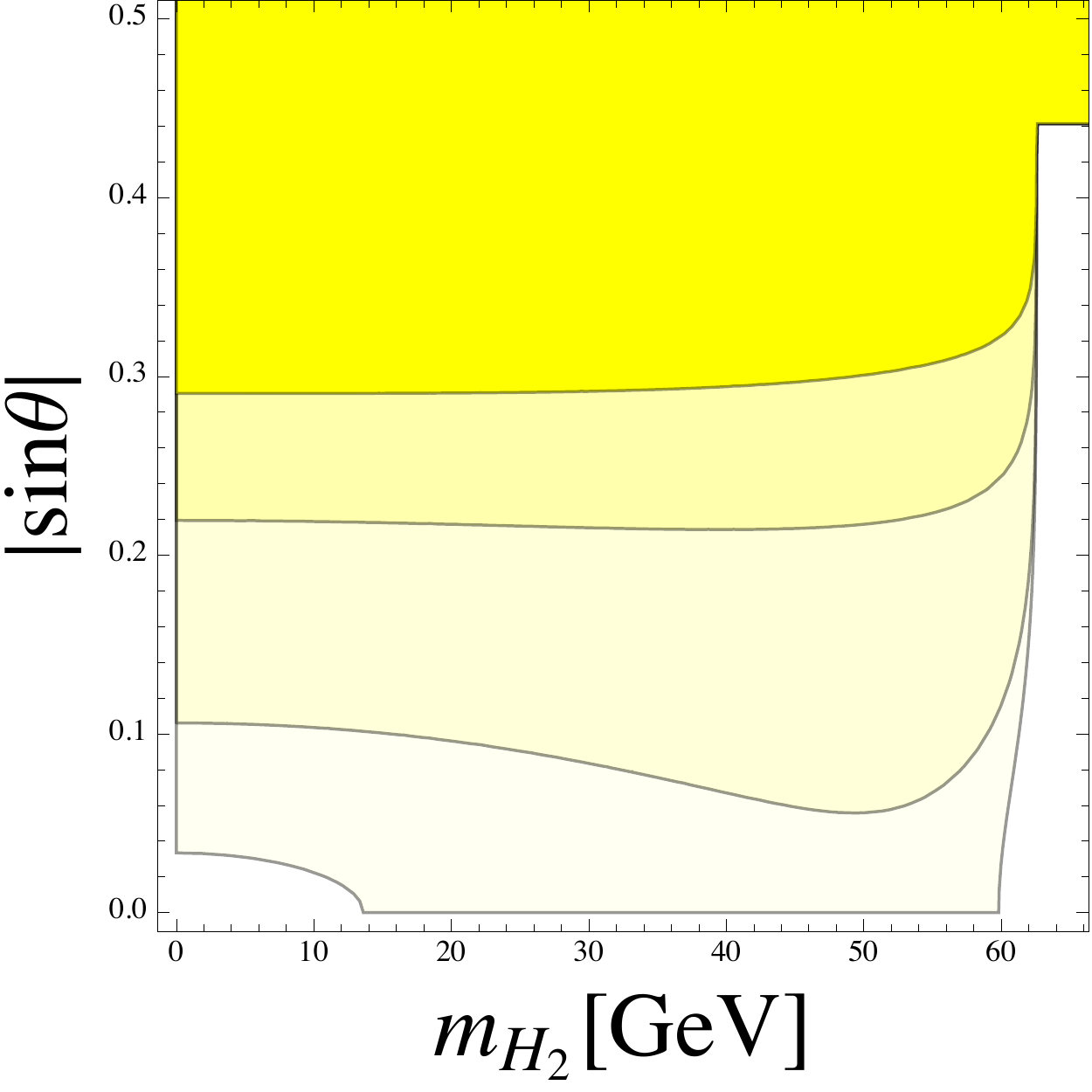} \quad 
 \includegraphics[width=0.47\textwidth]{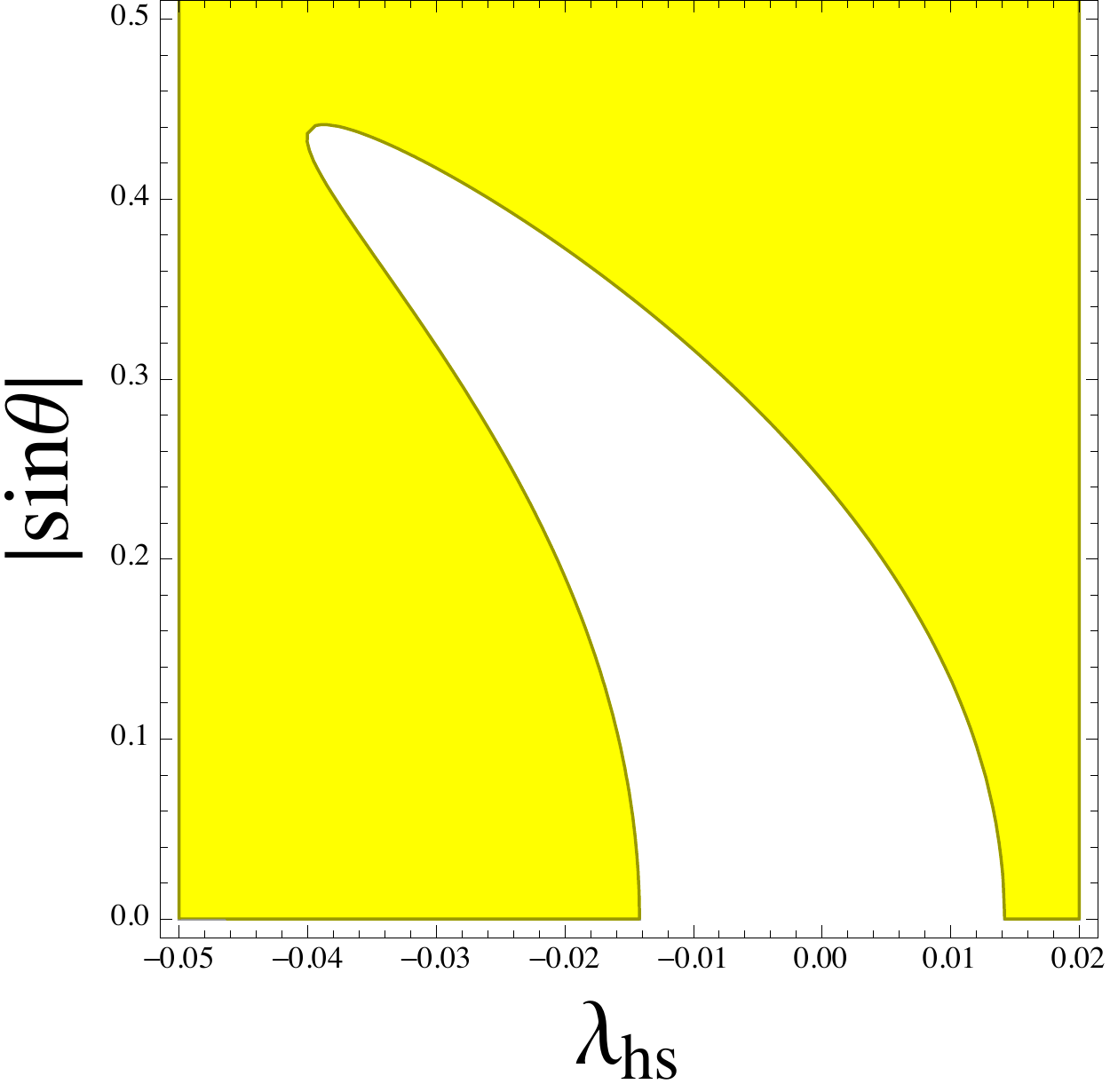}
%\vspace*{-2mm}
 \caption{
 {\em Left:} Regions of parameter space for 
$m_{H_2}<65$ GeV excluded at 95\% CL by the limits on the $H_1$ coupling. The excluded
areas (in yellow) correspond to 
$\lambda_{hs}= -0.011,0.0001,0.011,0.014$ (from the darkest to the palest). 
 {\em Right:} The excluded region (in yellow) for $m_{H_2}=20$ GeV. Inside
 the white region the $H_1 H_2^2$ coupling is very small. 
 }
\label{fig:h2h1h1}
\end{center}
\vspace*{-3mm}
\end{figure}

For $m_{H_2} < m_{H_1}/2$ the situation is more complicated because the $H_1 \to H_2 H_2$ decay channel opens up, 
leading to a decrease of the signal strength in all visible channels. 
This typically leads to stronger limits on $\sin \theta$, which are slightly dependent on $m_{H_2}$ and strongly on $\lambda_{hs}$. 
Representative examples of these constraints are shown in Fig.~\ref{fig:h2h1h1}.
A larger $\vert \lambda_{hs}\vert$ normally entails stronger limits and at some 
point almost the entire $m_{H_2}$--$\sin\theta$ plane gets excluded. Thus, at low 
$m_{H_2}$, the allowed $\lambda_{hs} $ is typically limited to the range
$\vert \lambda_{hs}\vert < 0.015$. 
Nevertheless, for a given $m_{H_2}$ and $\sin \theta$ one can always adjust a negative $\lambda_{hs}$ such that the $H_1 H_2^2$ coupling vanishes, in which case the limit is that of \eref{indirect_sinth}. 
The region excluded by $H_1$ coupling measurements is marked as the yellow area in \fref{all}.

%=========================================================================
\subsection{Limits from direct searches for a Higgs-like scalar}
%=========================================================================

Further constraints are provided by the LEP and LHC searches for a Higgs--like scalar. 
We take into account the following results: 
\bi
\item Searches for $H_2 \to \gamma \gamma$ in ATLAS~\cite{Aad:2014ioa} and CMS~\cite{Khachatryan:2014ira}. 
\item Combined limits from $H_2 \to ZZ $ and $H_2 \to WW $ searches in various final states in CMS~\cite{Khachatryan:2015cwa} (for $m_{H_2} \geq 145$ GeV) and searches for $H_2 \to ZZ $ in the $4 \ell$ channel in ATLAS~\cite{Aad:2014eva} and CMS~\cite{Chatrchyan:2013mxa} (for $m_{H_2} \leq 145$ GeV).
\item $H_2 \to H_1 H_1$ searches in CMS with the $2 b \, 2\gamma$~\cite{CMS:2014ipa} and $4b$~\cite{CMS:2014eda} final states, and in ATLAS with the $2 b \, 2\gamma$ final state~\cite{Aad:2014yja}.
\item LEP Higgs searches~\cite{Barate:2003sz} dominated by the $b \bar b$ decay channel. 
\item DELPHI search for a low mass Higgs in $Z$--decays~\cite{Abreu:1990bq}.
\item $b$--physics constraints on a low mass Higgs~\cite{Aaij:2012vr,Lees:2012iw}.
\ei 

The parameter space excluded by these searches is shown as the red area in \fref{all}. 
At very low masses, $m_{H_2} < 5$ GeV, the strongest limits come from $B \rightarrow 
K \ell \ell $ decays~\cite{Aaij:2012vr}. The resulting constraint 
$\sin\theta < 10^{-2} \ldots 10^{-3}$ can be 
extracted from the analysis of Ref.~\cite{Schmidt-Hoberg:2013hba}.
Between 5 GeV and 12 GeV, the bound $\sin\theta \lesssim 0.5$ is imposed by the radiative 
$\Upsilon$ decays~\cite{Lees:2012iw} and the DELPHI searches for a light Higgs in 
$Z$--decays~\cite{Abreu:1990bq}. Above this mass window up to about
115 GeV, LEP Higgs searches~\cite{Barate:2003sz} become relevant. The resulting bound 
on $\sin\theta$ is about $1 \times 10^{-1}$ to few$ \times 10^{-1}$ depending
on the exact $H_2$ mass.
The region between 120 and 130 GeV remains poorly constrained due to the presence
of the SM--like Higgs,\footnote{This allows for an almost degenerate second Higgs with 
a large mixing between the two~\cite{Heikinheimo:2013cua}.
} 
whereas a strip just below and above it is constrained
through the diphoton channel searches~\cite{Aad:2014ioa,Khachatryan:2014ira} although the bound is still looser than the indirect one from the $H_1$ coupling measurements. 
Above $\sim 135$~GeV the limits are dominated by $H_2 \to ZZ $ and $H_2 \to WW $ searches~\cite{Khachatryan:2015cwa,Aad:2014eva,Chatrchyan:2013mxa}.
This imposes $\sin\theta < 0.3 \ldots 0.4$ in a wide range of masses up to about 500 GeV, above which the indirect bounds are consistently stronger.

Concerning the $\lambda_{hs}$--dependence of the exclusion limits, let us note that
the limits on $(m_{H_2}, \sin\theta)$ can be much stronger for a given $\lambda_{hs}$.
In particular, for $2 m_{H_2} \leq m_{H_1}$ the $H_1 \rightarrow H_2 H_2$ decay
would dilute the Higgs signal strength as explained in the previous subsection. 
Therefore, in \fref{all} we marginalise over $\lambda_{hs}$ in this mass region.
For $m_{H_2} > 2 m_{H_1}$ the limits also depend on $\lambda_{hs}$: the larger it is, the more suppressed are the $H_2 \to ZZ $ and $H_2 \to WW $ channels, and the more enhanced is the $H_2 \to H_1 H_1$ decay. 
However, this effect is non-negligible only for $\lambda_{hs} \gtrsim 1$; for smaller values the limits depend very little on the precise value of $\lambda_{hs}$. 
The constraints from $H_2 \to H_1 H_1$ are more important than those from $H_2 \to ZZ $ and $H_2 \to WW $ only for $\lambda_{hs} \gtrsim 2$; for no perturbative value of $\lambda_{hs}$ are these limits stronger than the indirect ones from the $H_1$ coupling measurements.

%=========================================================================
\subsection{Combined experimental constraints vs. vacuum stability}
%=========================================================================

Combining the bounds from direct searches, precision tests, and $H_1$ coupling measurements and imposing them on the parameter space favoured by the stability
considerations in \fref{all} (green), we find that for $m_{H_2} \gtrsim 200$ GeV all of
the constraints are compatible. The entire stability--favoured region above $\sim 350$ GeV is unconstrained, whereas between 200 and 350 GeV there are pockets of allowed 
parameter space with $\sin\theta $ between 0.2 and 0.4.

The favoured region can be probed further by measuring the Higgs signal strength with higher precision as well as by searching for the decay $H_2 \rightarrow H_1 H_1$.

\begin{figure}[h]
 \begin{center}
 \includegraphics[width=0.47\textwidth]{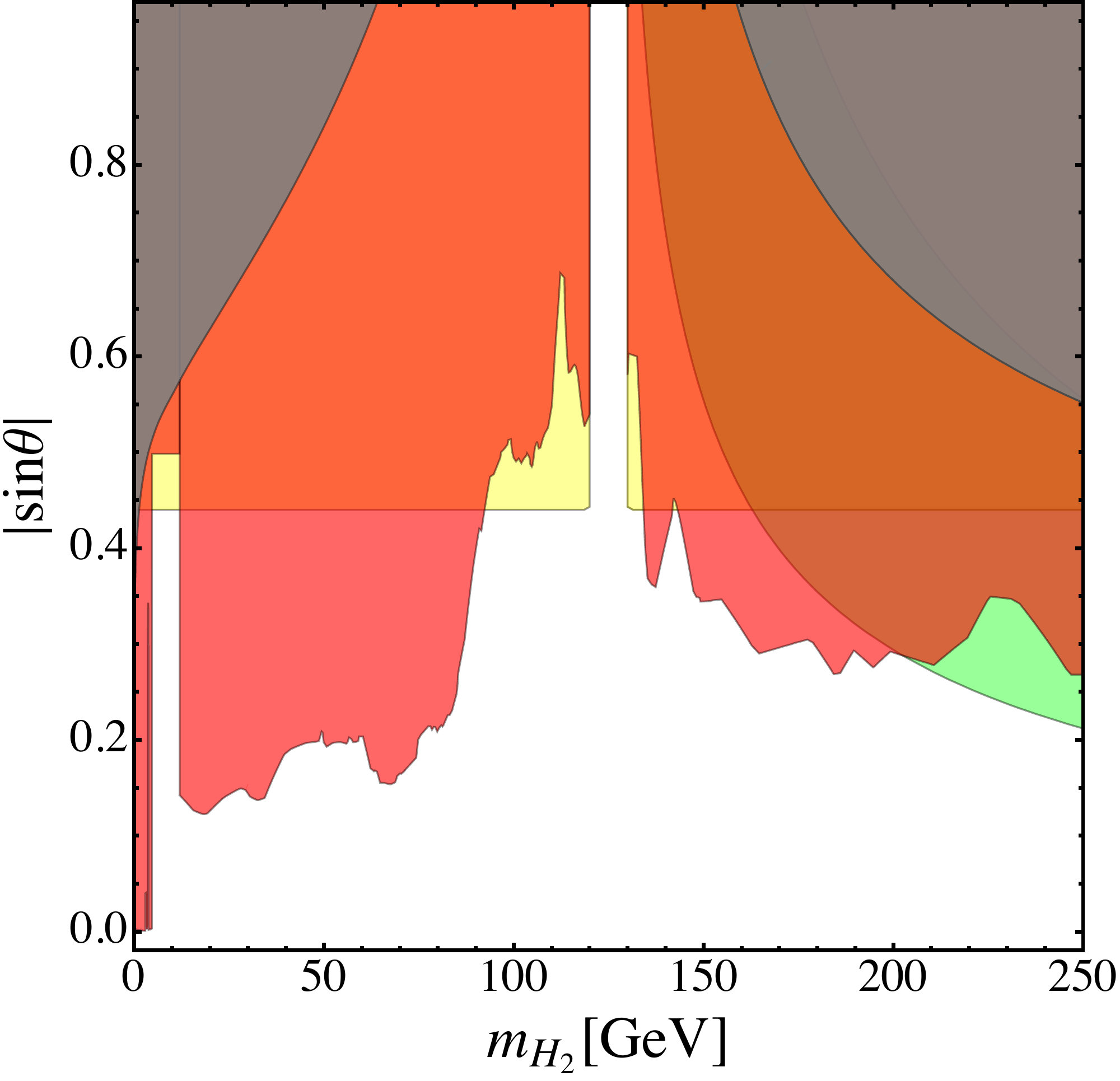} \quad 
 \includegraphics[width=0.47\textwidth]{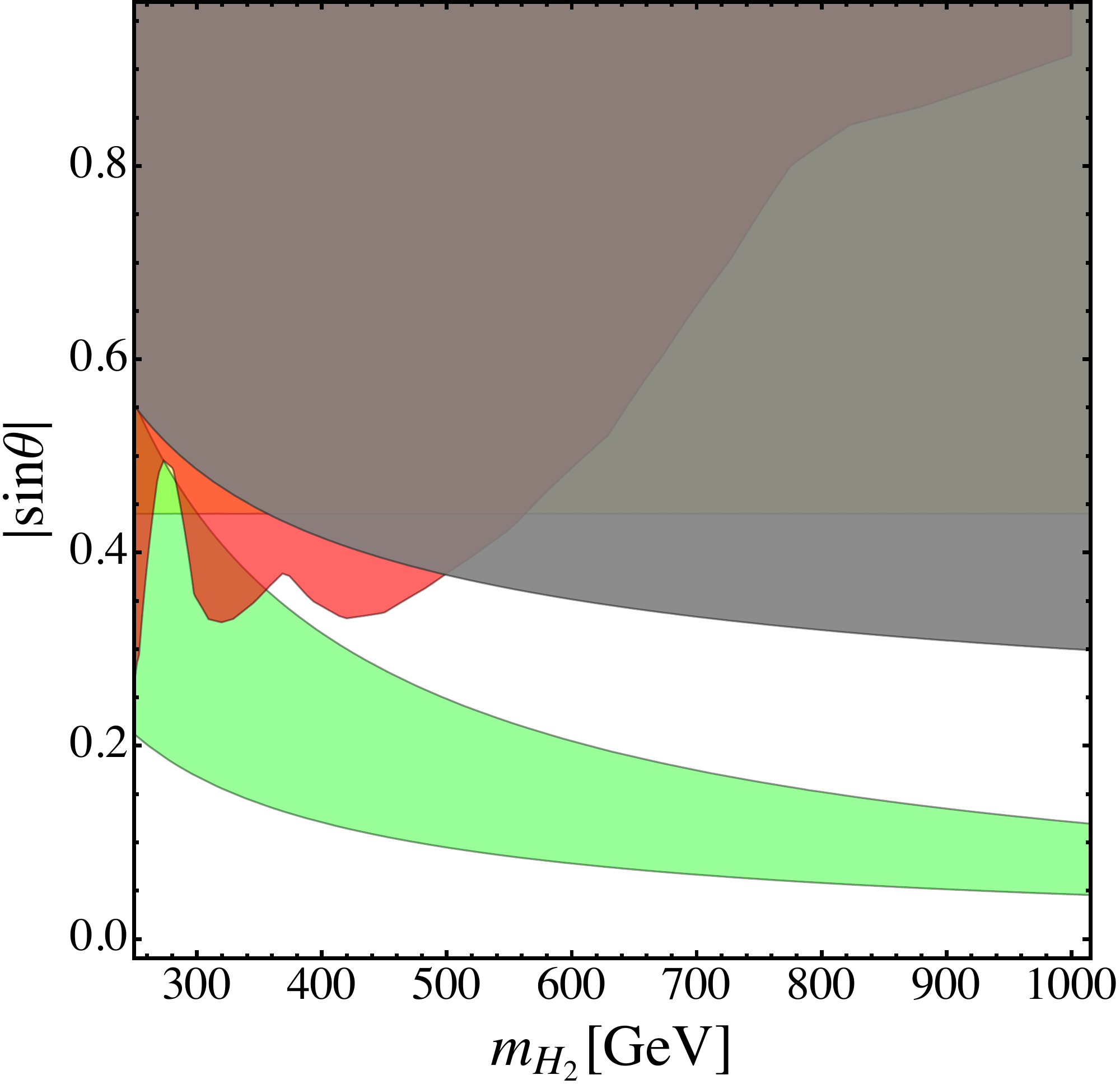}
%\vspace*{-2mm}
 \caption{ {\em Left:} 
 Parameter space (for $m_{H_2} \leq 2m_{H_1}$) excluded at 95 \% CL by 
 direct searches (red), precision tests (gray), and $H_1$ couplings measurements (yellow). For $m_{H_2} < m_{H_1}/2$, the limit from the $H_1$ couplings is marginalised over $\lambda_{hs}$, otherwise it does not depend on $\lambda_{hs}$. 
The green region is preferred by stability of the scalar potential up to the Planck scale at $\lambda_{hs} = 0.01$; 
for other $\lambda_{hs} $, it is either very similar or smaller and contained within the green region. 
 {\em Right:} Same for $m_{H_2} > 2 m_{H_1}$. 
 }
\label{fig:all}
\end{center}
\vspace*{-3mm}
\end{figure}

%=========================================================================
%=========================================================================
\section{Prospects for observing $H_2 \to H_1 H_1$ at LHC-13}
%=========================================================================
%=========================================================================

For $m_{H_2} > 250$ GeV, the decay $H_2 \rightarrow H_1 H_1$ is allowed kinematically. It is an important channel for studying properties of $H_2$, which allows for a reconstruction of $\lambda_{hs}$
\cite{Englert:2011yb}. The rate of $H_2 \rightarrow H_1 H_1$ depends on $\sin\theta$, $m_{H_2}$ and also
$\lambda_{hs}$, cf.~Eq.~(\ref{gamma211}). While the first two parameters can be fixed using the SM--like decay modes of $H_2$,
determination of $\lambda_{hs}$ requires an additional channel such as $H_2 \rightarrow H_1 H_1$.

\begin{figure}[h]
\begin{center}
 \includegraphics[width=0.38\textwidth]{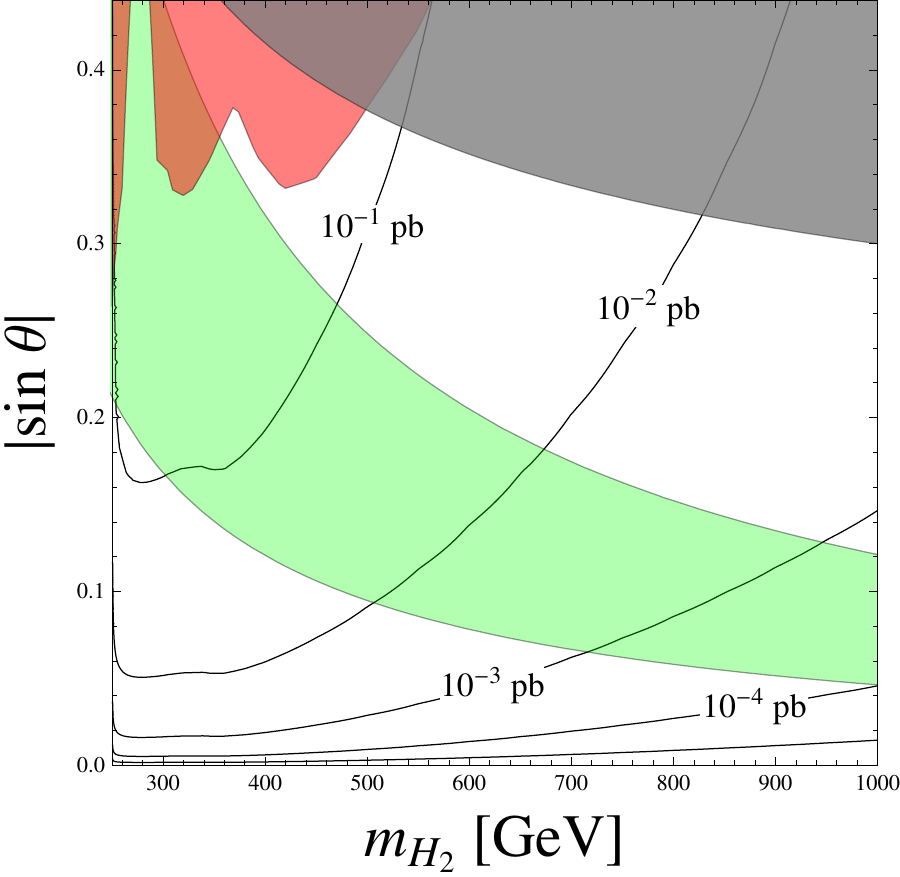}
\quad 
\includegraphics[width=0.575\textwidth]{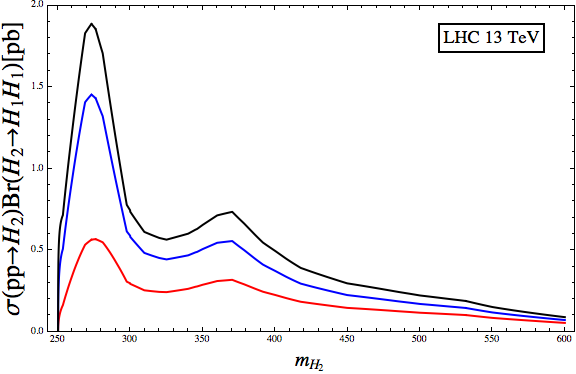}
\vspace*{-2mm}
 \caption{
 {\em Left:} 
 $\sigma(pp \rightarrow H_2)~$BR$(H_2 \rightarrow H_1 H_1)$ at LHC-13 for $\lambda_{hs}=0.01$ in the $\sin \theta$-$m_{H_2}$ plane.
 {\em Right:} 
$\sigma(pp \rightarrow H_2)~$BR$(H_2 \rightarrow H_1 H_1)$ at LHC-13 for maximal 
allowed values of $\sin \theta$, 
with $\lambda_{hs}=0.01$ (bottom), $\lambda_{hs}=1$ (middle), $\lambda_{hs}=2$ (top). $m_{H_2}$ is in GeV.
 }
\label{fig:rate}
\end{center}
\vspace*{-3mm}
\end{figure}

The left panel of Fig.~\ref{fig:rate} displays contours of equal $\sigma(pp \rightarrow H_2)~$BR$(H_2 \rightarrow H_1 H_1)$ in the $\sin \theta - m_{H_2}$ plane, while the right panel shows the maximal production rate $\sigma(pp \rightarrow H_2)~$BR$(H_2 \rightarrow H_1 H_1)$ at LHC-13 consistent with all the experimental constraints. 
The different curves in the right panel correspond to different $\lambda_{hs}$.
At fixed $\lambda_{hs}$, the rate is restricted by the bound on $\sin\theta$ which is mostly due to the LHC constraints for $m_{H_2} < 500$ GeV and to the electroweak constraints for $m_{H_2} > 500$ GeV.
The rate also increases with $\lambda_{hs}$, which we take to be $0.01,1,2$ in the plot. In all of these cases,
$\sigma(pp \rightarrow H_2)~$BR$(H_2 \rightarrow H_1 H_1)$ is in the picobarn range for $m_{H_2}$ up to about 400 GeV.
This makes the prospects for detecting $H_2 \rightarrow H_1 H_1$ at LHC-13 quite good, at least for a relatively light $H_2$.

Imposing the extra stability/perturbativity constraint up to $m_{P}$, we find reduction of the maximal rate
for $m_{H_2}$ above around 350 GeV. This theoretical constraint becomes the strongest bound on the model, with
the preferred parameter space being difficult to probe experimentally. For light $H_2$ however, the main constraints are due to the LHC heavy Higgs searches which allow for a substantial rate
$\sigma(pp \rightarrow H_2)~$BR$(H_2 \rightarrow H_1 H_1)$.

%=========================================================================
%=========================================================================
\section{Summary and conclusions}
%=========================================================================
%=========================================================================

We have analysed constraints on the two scalar states of the simplest Higgs portal model. One of them is identified with the 125 GeV Higgs--like boson observed at the LHC.
The other boson's mass is allowed to be in a wide range down to about 5 GeV, below which the constraints on the mixing angle become severe. 
Above $\sim 90~\GeV$, a substantial mixing between the Higgs and the singlet, $\sin\theta \sim 0.3 \ldots 0.4$, is consistent
with the data. 

Stability of the scalar potential can be improved over that of the SM if the 
state $H_2$ is sufficiently heavy, above about 200 GeV. For a range of $\sin\theta$
consistent with the electroweak precision measurements and the LHC data,
the electroweak vacuum is stable and the model is perturbative up to the Planck scale.
The required mixing angle is of order $10^{-1}$ for $m_{H_2}$ up to 1 TeV.
 
In the allowed parameter space, the decay $H_2 \rightarrow H_1 H_1$ can be quite efficient such that the $H_1$ pair production rate at LHC-13 is at the picobarn level.
This applies to a relatively light $H_2$ up to about 400 GeV, with the rate quickly falling off above 500 GeV or so. Apart from the search for a new resonance, the Higgs portal can be efficiently constrained by further improvement of the Higgs coupling measurements.

\vspace{10pt}

{\bf Acknowledgements.} This work was supported in part by the Academy of Finland project ``The Higgs boson and the Cosmos''.
AF~is supported by the ERC Advanced Grant Higgs@LHC. 

\vspace{10pt}

{\it N.B.} At the time of completion of this work, we became aware of the preprints arXiv:1501.02234 (Ref.~\cite{Pruna:2013bma}) and arXiv:1501.03799 (Ref.~\cite{Martin-Lozano:2015dja}) which have some overlap with our study.

 %%%%%%%%%%%%%%%%%%%%%%%%%%%%%%%%%%%%%%%%%%%%
%%%%%%%%%%%%%%%%%%%%%%%%%%%%%%%%%%%%%%%%%%%%
{}

\end{document}